\documentclass[a4paper,11pt]{article}
\usepackage{pos}
\usepackage{wrapfig}
\usepackage{subfig}

\title{Episodic ejection from a low-mass young stellar object traced by H$_2$O masers}

\author*[a,b]{Zs. M. Szabó}
\author[c]{O. Bayandina}
\author[a,b,d]{Á. Kóspál}
\author[a,b]{P. Ábrahám}
\author[b]{K. \'E. Gabányi}
\author[a]{Zs. Nagy}
\author[b]{V. L. Tóth}
\author[e]{S. P. van den Heever}

\affiliation[a]{Konkoly Observatory,\\
  Konkoly Thege Miklós út 15-17, Budapest, Hungary}
  
\affiliation[b]{E\"otv\"os Lor\'and University, Department of Astronomy, P\'azm\'any P\'eter s\'et\'any 1/A, 1117 Budapest, Hungary}

\affiliation[c]{Joint Institute for VLBI ERIC,\\
  Oude Hoogeveensedijk 4, Dwingeloo, The Netherlands}

\affiliation[d]{Max Planck Institute for Astronomy,\\
  Königstuhl 17, Heidelberg, Germany}
  
\affiliation[e]{South African Radio Astronomy Observatory,\\
  Broederstroom Road, Hartebeesthoek, South Africa}

\emailAdd{szabo.zsofia@csfk.org}

\abstract{We present the project of a VLBI study of the 22 GHz H$_2$O maser in a prototypical low-mass
protostellar system IRAS 16293-2422. The observation was conducted to characterise the cause of the newly discovered
enhanced maser activity in the source and to study the source's ejection behaviour as traced by maser
emission. Single-dish monitoring and analysis of archival data indicate that the activity of the
H$_2$O maser in IRAS 16293-2422 has a cyclic character and traces episodic ejection events in
the source. A new maser flare was recently discovered in a spectral feature that has never
shown such a significant increase in flux density before. The flare of this feature seems to
indicate the beginning of a new cycle of activity.}

\FullConference{%
  *** European VLBI Network Mini-Symposium and Users' Meeting (EVN2021) ***\\
  *** 12-14 July, 2021 ***\\
  *** Online ***
}

\begin{document}
\maketitle

\section{Masers to trace star formation}

 The life cycles of stars follow patterns based mostly on their initial mass. 
Comparison of the physical parameters of protostar envelopes of different masses hints that the transition between them seems to be smooth, and the formation processes and triggers are similar \citep{Crimier2010}.
 Disk-mediated accretion accompanied by episodic accretion bursts is thought to be a common mechanism of star formation across the entire stellar mass spectrum. 
 However, the accretion process itself is poorly understood due to scarce observational evidence.

Much progress in the study of high-mass star formation has been made recently with the study of masers, which have proven to be a powerful tool for locating massive young stellar objects (YSOs) undergoing accretion events (see \citep{Burns2020}, and other publications of the \href{https://masermonitoring.org}{Maser Monitoring Organisation (M2O)} - a global co-operative of maser monitoring programs). 

During accretion events in massive YSOs, multiple maser species and transitions flare (e.g. \citep{Brogan2019}).
Maser action occurs only over certain ranges of physical conditions (e.g. \citep{Ellingsen2007}), hence the spatial distribution of masers can reveal the temperature, density and radiation enhancements in the region, while the kinematics of the maser spots can indicate gas motions. 

According to theory and observation \citep{Caratti2015}, there is a correlation between ejection and accretion rates, so major outflow events can be linked to major accretion events (bursts).
Under this hypothesis the accretion history of a star can be inferred by its ejection history traced by symmetric pairs of ejection bow shocks at ever increasing distances from the central
object \citep{Burns2017}. 
The morphology of the jets/outflows traced by H$_2$O masers may indicate timescales of the accretion process, reveal the collimation properties of the jet and the density of the ambient material.



However, while the study of maser emission has shown excellent results for massive YSOs, adaptation of the approach for low-mass protostars is challenging. Low-mass YSOs show much less variety of associated maser species, and the detected masers are highly variable and (in most cases) weak.

 \section{IRAS 16293-2422}
 
IRAS 16293-2422 (IRAS 16293, hereafter) is a low-mass protostellar system, showing bright and active H$_2$O maser emission.
The source is known for its very rich chemistry, several complex molecules have been discovered in it for the first time ever (e.g. \citep{jorgensen2012, Fayolle2017}).

The source is a binary system, usually referred to as sources A and B 
\citep{Mundy1992}.
The sources A and B have properties of Class 0 objects.
Component B appears to be a single source, while component A was resolved into two subcomponents: A1 (an ionized region associated with an outflow) and A2 (a protostar powering a radio jet) \citep{Chandler2005, Loinard2007}. 
An almost edge-on rotating disk around source A and signs of infall around source B were found in \citep{Pineda2012}. The latest multi-epoch continuum VLA observations of the system confirm that A2 is a protostar driving episodic mass ejections \citep{Hernandez2019}.



The source A2 shows highly variable, but bright 22 GHz H$_2$O maser with the flux density of tens of kJy during flares and no less than 100 Jy in a stable state \citep{Colom2016}.
The flaring features appear alternately at blueshifted and redshifted
velocities with respect to the cloud velocity of~$\sim$4~km~s$^{-1}$,
typically in the velocity range of~$-$5~--~10~km~s$^{-1}$.  Recently a new maser flare of tens kJy was detected in a blueshifted feature at the velocity of $-$1.5 km~s$^{-1}$; this is the brightest flare ever detected in this spectral feature.


VLBI studies of the spatial distribution of the 22 GHz H$_2$O maser emission in the source showed that there are two separate clusters: maser spots with blueshifted velocities are found to the north, and redshifted spots are found to the south \citep{Imai1999, Imai2007, Dzib2018}. Most of the papers propose to associate the detected H$_2$O maser clusters with outflows (e.g. \citep{Imai2007}). Single-dish monitoring of maser emission in the source suggests that the H$_2$O maser flares are most probably linked to motions of shocked gas \citep{Colom2016}.

\begin{figure}[h]
\centering
\includegraphics[width=1\linewidth]{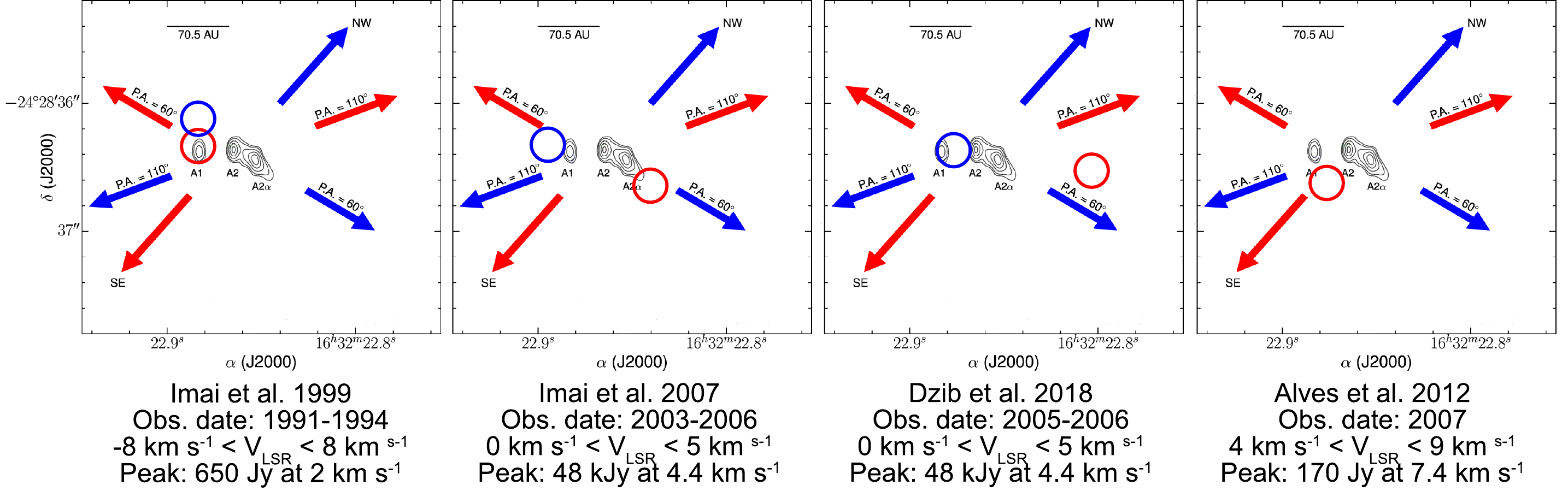}
\caption{\footnotesize{The evolution of the 22 GHz H$_2$O masers' special distribution in IRAS 16293. Background image is from \citep{Dzib2018}: A1 and A2 - the sources composing the system; blue and red arrows mark the directions of the three outflows in the system. The blue and red circles mark the average positions of blueshifted and redshifted H$_2$O maser features, respectively. For each observation epoch (see the reference under each map), the velocity range of the detected H$_2$O maser features with the peak flux and velocity is indicated.}}
\end{figure}

Our analysis of the VLBI data available in literature indicates that H$_2$O masers in IRAS 16293 could trace shocks excited by a precessing outflow system (see Fig.1). Although the general segmentation of maser emission into two spatial clusters persists from epoch to epoch, the distance and alignment between them vary. 
The elongated morphology and bipolar outward motion of water masers suggest that they are associated
with an ejection event from the YSO. Particularly noteworthy is the fact that the distance between clusters increases.  
There is a correlation between kinematic age and outflow length (the angular separation between the lobes) of H$_2$O maser jets: very young bipolar H$_2$O maser jets/outflows are very compact. 

Note that observation of \citep{Alves2012} was made during a quiescent state of the maser and blueshifted features were not detected. Monitoring of \citep{Colom2016} indicated that the activity of the H$_2$O maser in IRAS 16293 has a cyclic character, and the period of 2006-2008 was one of the longest maser emission minima. Thus, it is possible that the shocks traced by the H2O maser and presented in \citep{Imai1999, Imai2007} and \citep{Dzib2018} no longer exist, and with VLBI observation of the maser emission associated with the new flare, we can catch the launch of young ejection bow shocks.

 \section{The EVN observation}
 
 The EVN observation of the source was conducted on March 11, 2021. The telescope array consisted of 19 antennas, including the longest NS baselines to the 26-m Hartebeesthoek telescope and the longest EW baselines to the KVN telescopes (see the obtained UV-coverage in Fig.2). The achieved spatial resolution was $\sim$1~mas. 
 
 By the date of the EVN observation, the flare flux density of the source had decreased from tens of kJy to $\sim$2~kJy (Fig.3). Nevertheless, such a high brightness of the source made it possible to study in detail the spatial distribution of the maser features. All of the spectral features were detected on the EVN baselines with a high signal-to-noise ratio.
 A detailed analysis of the obtained data will be presented in subsequent publications.
 

\begin{figure}[h]
\centering
\begin{minipage}{.45\textwidth}
\centering
  \includegraphics[width=0.8\linewidth]{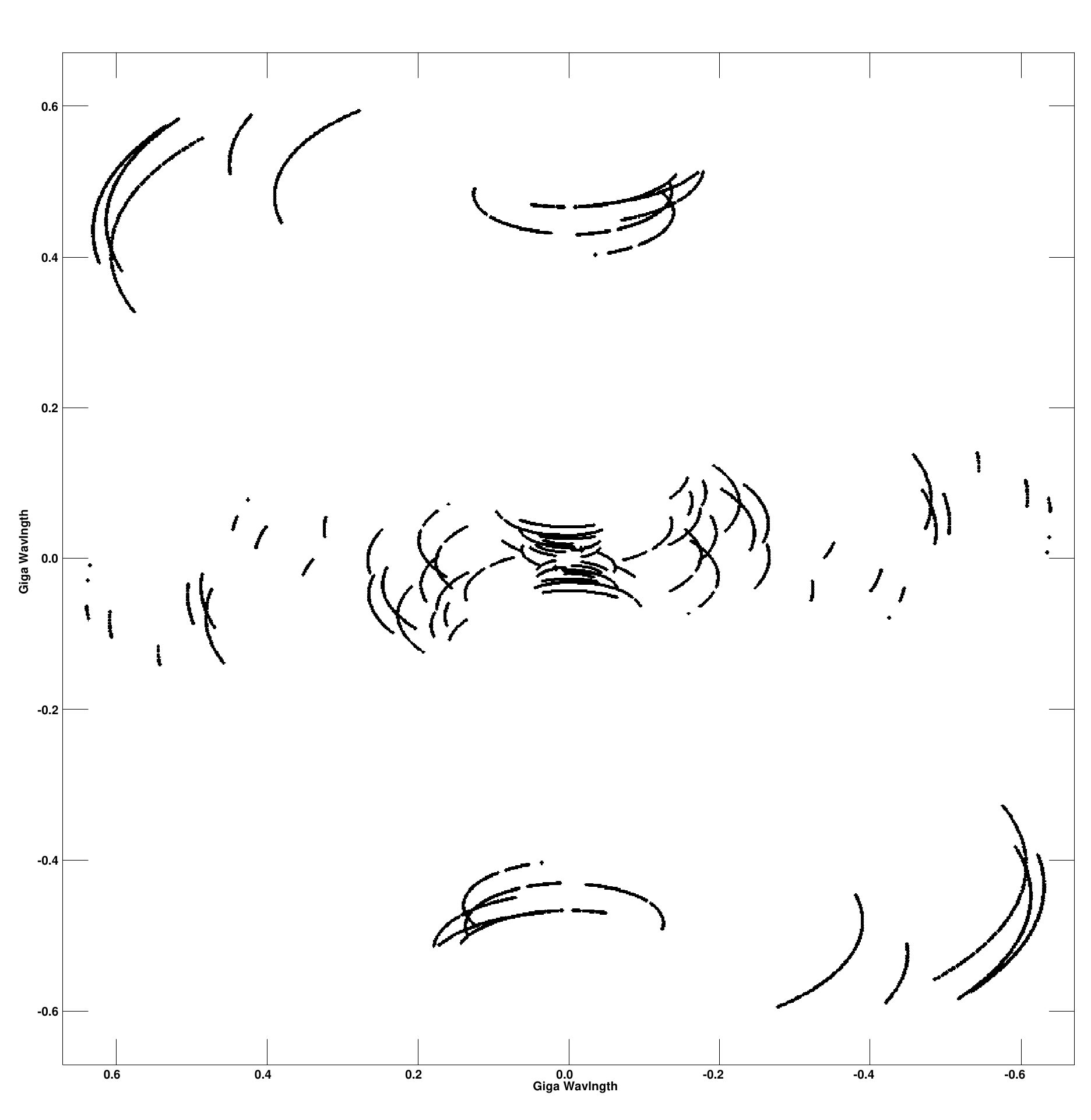}
 \captionof{figure}{\footnotesize{The UV-coverage obtained  for the target source IRAS 16293 in the EVN observation on March 11, 2021.}}
\end{minipage}
\hspace{5mm}
\begin{minipage}{.45\textwidth}
\centering
\includegraphics[width=1.1\linewidth]{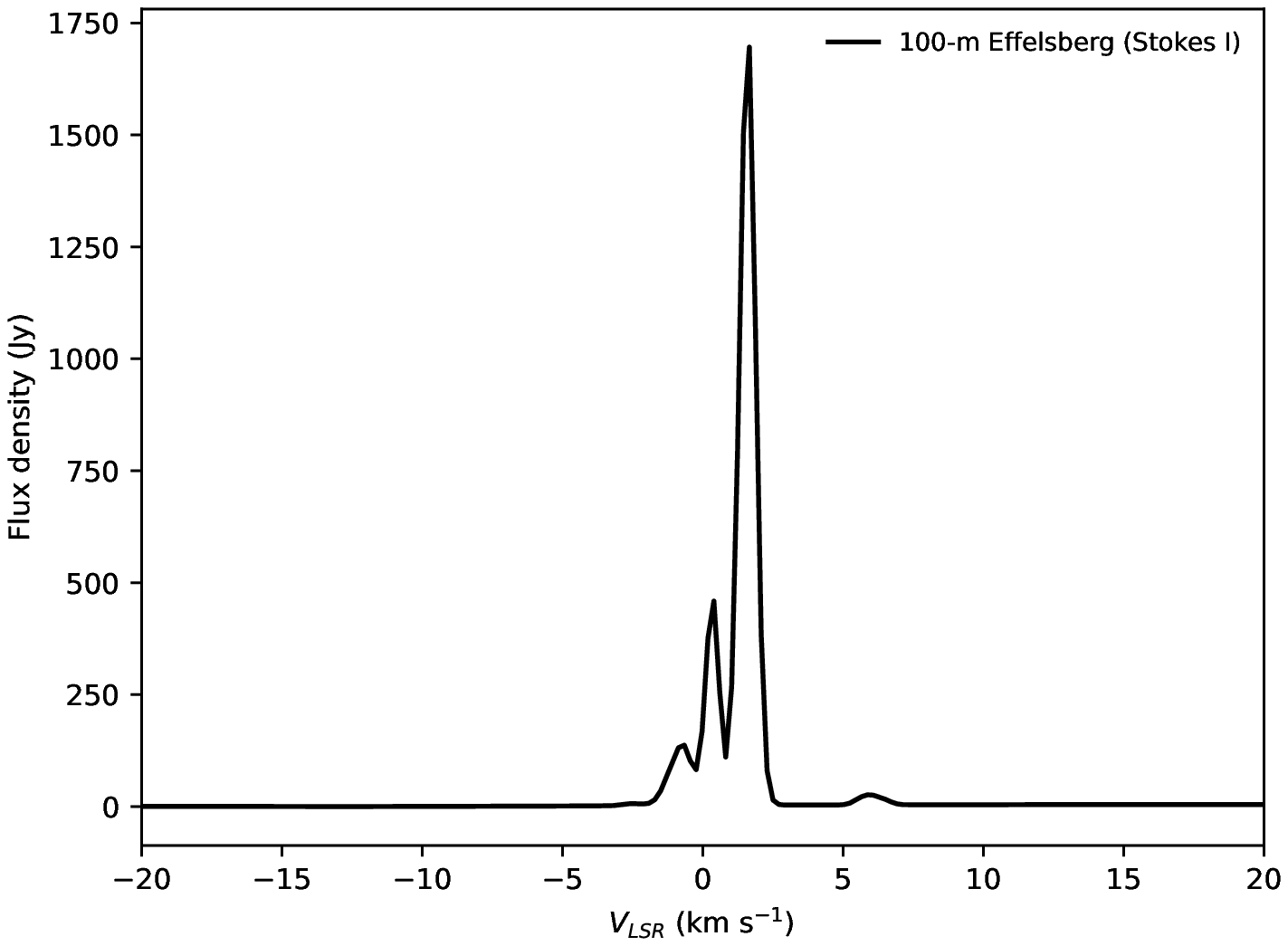}
\captionof{figure}{\footnotesize{\footnotesize{The auto-correlation spectrum of the flaring 22 GHz H$_2$O maser in IRAS 16293 obtained with the 100-m Effelsberg telescope.}}}
\end{minipage}%
\end{figure}



\end{document}